# Human search in a fitness landscape: How to assess the difficulty of a search problem

Oana Vuculescu[1], Mads Kock Pedersen[2*], Jacob F. Sherson[2], Carsten Bergenholtz[1]

[1] Department of Management, Aarhus University, Aarhus, Denmark, [2] ScienceAtHome, Department of Physics and Astronomy, Aarhus University, Aarhus, Denmark
[*]corresponding author: madskock@phys.au.dk

Abstract
Computational modeling is widely used to study how humans and organizations search and solve problems in fields such as economics, management, cultural evolution and computer science. We argue that current computational modeling research on human problem-solving needs to address several fundamental issues in order to generate more meaningful and falsifiable contributions. Based on comparative simulations and a new type of visualization of how to assess the nature of the fitness landscape, we address two key assumptions that approaches such as the NK framework rely on: that the NK captures the continuum of the complexity of empirical fitness landscapes, and that search behavior is a distinct component, independent from the topology of the fitness landscape. We show the limitations of the most common approach to conceptualize how complex, or rugged, a landscape is, as well as how the nature of the fitness landscape is fundamentally intertwined with search behavior. Finally, we outline broader implications for how to simulate problem-solving.

# Introduction

*"Solving a problem simply means representing it so as to make the solution transparent"* [1]

There is a long tradition of studying how to search for solutions to 'hard' problems, i.e. problems where it is computationally impossible or merely too expensive to list and test all possible solutions [2, 3]. The prevalent way of addressing individual or organizational search behavior and how to conceptualize the space of solutions stems from early work on population genetics, namely the fitness landscape model [4]. By focusing on fitness interactions between genes, Wright's framework allows for a link between low-level properties of genes and the high-level patterns of the dynamics of evolution [5]. The model's most famous extension, the NK model [6], explicitly models adaptive evolution as a "search in protein space" [6] which tries to find a maximum point for a chosen fitness function. This approach has grown outside the boundaries of population genetics literature and inspired a series of scholars from computer science [7], organizational theory [8, 9, 10], economics [11], cultural evolution [12], and physics [13, 14] to computationally model complex, adaptive systems.

How can problem-solving be modeled in this framework? Imagine trying to solve an innovation problem, for instance designing a new educational app. The app will likely be based on predefined libraries, which constitute interconnected modules; changing something in one module, might influence the functionality of another module. The extent of this interdependence will differ from environment to environment, and influence how one should search for the good design. Sometimes changing one small element at a time ('local-search') might be efficient, while in other environments the level of interdependency might make this approach inefficient.

Levinthal [15] introduced the NK model to the social science literature in order to facilitate formal modeling and simulation of how the level of interdependence of organizational activities affects its long-term chances of finding the optimal solution to a problem and thus survive in a competitive environment. By making explicit assumptions about individual or organizational behavior and the environment in which the agent evolves, researchers could now simulate how such agents adapt over time. As in the study of genes in biology, this allows one to map the complex dynamics of agents being embedded in and adapting to the competitive environment [8, 16-20].

In the NK model all the positions $x$ in the fitness landscape belongs to a $N$-dimensional space where the value along each of the dimensions can take the values 0 or 1 i.e.

$$x \in \{0,1\}^N,$$

representing which attributes of genes are active or not. The neighborhood size parameter $K$ determines the level of interdependency among the dimensions when evaluating the fitness value $f(x)$. Thus, in order to evaluate the fitness value for $x$ you compute the value of the $N \cdot 2^{K+1}$ given local fitness functions,

$$f_i(x) = f_i(x_i; x_1^i, \ldots, x_K^i),$$

where $x_1^i, \ldots, x_K^i$ are the values of the $K$ successive neighbors to the $i$'th dimension. The neighbors are taken in a cyclic manner e.g for $K = 3$ neighbors to $x_{N-1}$ would be $x_N$, $x_1$, and $x_2$. The fitness value of $x$ is then the average of the local fitness functions.

$$f(x) = \frac{1}{N} \sum_{i=1}^{N} f_i(x)$$

An example of an $N=3$, $K=1$ landscape with the local fitness functions could be

$$f_1(00) = 0, f_1(01) = 0.64, f_1(10) = 1, \text{ and } f_1(11) = 0.75,$$

$$f_2(00) = 0.80, f_2(01) = 0.14, f_2(10) = 0.42, \text{ and } f_2(11) = 0.91,$$

$$f_3(00) = 0.79, f_3(01) = 0.95, f_3(10) = 66, \text{ and } f_3(11) = 0.04.$$

The fitness value of $x = 110$ would be

$$f(110) = \frac{1}{3}(f_1(11) + f_2(10) + f_3(01)) = \frac{1}{3}(0.75 + 0.42 + 0.95) = 0.71.$$

For further details on how the NK landscape is computed see Ganco and Hoetker, 2009 [10].

The original NK model [6] assumes a landscape where one can modify the complexity by varying the interdependence of elements (the $K$ of the NK). This ability to vary the interdependence is the model's great strength, compared to less flexible simulation frameworks such as armed bandits [9]. Furthermore, the model also assumes that the searcher primarily engages in local-search akin to the one-bit-flip mutations of a gene. We want to address these assumptions in a social context and offer a technical, comparative analysis of search and the empirical landscapes we intend to model. First, we discuss how social science studies usually conceptualize how complex (i.e. difficult to search) a given landscape is. However, we show that the commonly used measure for how rugged a landscape is does not capture how social science landscapes might be ordered by a hierarchy [21] or how neutrality influences ruggedness [18]. Second, we provide a novel (to the social science literature) type of visualization that maps how different search strategies actually 'generate' different landscapes, rather than merely constituting search in an a priori given space. The conceptualization of the fitness landscape is thus not independent of assumptions about search behavior. This interdependence was not an issue in the original NK model since it assumed a particular search strategy of genes engaging in local-search with occasional random jumps [6]. Yet, if one changes the parameters of search behavior, the assessment of the difficulty of a fitness landscape should be able to take the impact of changing parameters into account. Overall, we argue that in line with recent trends in biology and computer science, we need to move beyond a simple numerical classification of the ruggedness of a fitness landscape, since such numerical classifications can not take all relevant landscape features (such as neutrality and deceptiveness) into account. We, therefore, contend that future research should rely on an overall categorization of the type of fitness landscape at hand, which could then shed light on which search algorithms to apply in a given situation [22]. This also points to a need to engage in further and wider comparisons of the NK fitness landscape features with characteristics found in the empirical world [23], both in terms of the task structure as well as the search behavior encountered in the empirical world.

## Search in NK fitness landscapes

There are two main elements in the fitness landscape model that need to be specified for problem-solving processes to be captured [24]: the task structure (i.e. the problem that is to be solved) and the search behavior (i.e. how problem-solving unfolds). In each step of the simulation, the agent follows pre-specified search rules, in an attempt to find the optimal solution, i.e. the peak of the fitness landscape. The landscape is a mapping between solutions and fitness values that takes into account the connectivity between solutions in the search space. But, in order to define this connectivity, we need to specify a distance metric that informs how agents can move between different positions in the search space. Different distance metrics lead to different landscape structures, a topic we will return to later.

### K/N ratios

Social science studies relying on the NK model follow on the path proposed by Kauffman [6] and study how the attributes of the search space influence the propensity of finding the optimal solution. Reflecting on the nature of genetic mutations, the default search strategy is generally local-search, i.e. a one-bit-flip hill-climber [15, 25]. In a landscape consisting of many local peaks, a local-search strategy is likely to lead the agent to become stuck before reaching the global peak, which is why other search strategies are often integrated, e.g. random jumps or directed search based on a cognitive representation of the landscape [26, 27, 28]. The K/N ratio is an attempt to describe how rugged a landscape is. The more bits are interconnected compared to the size of the bitstring, the more interdependencies have to be taken into account when evaluating the fitness value. Part of the NK model's popularity in organizational literature is due to allowing the investigation of different problem difficulties [29], via the K/N ratio, or the level of epistatic interactions [30]. Epistasis is equivalent to the non-linearity of a problem or how well a problem can be decomposed into sub-problems [7, 31]. Consider a N=4, K=1 example. A solver with $\{x = (0,0,0,1), f(x) = 0.56\}$, might move to $\{x_1 = (0,0,0,0), f(x_1) = 0.58\}$, even though $\{x_2 = (0,0,1,1), f(x_2) = 0.72\}$. The fact that the optimal setting for the fourth allele is $\{1\}$, is obscured by the epistatic interaction with its neighbor on the third position. This is also why search on such a rugged landscape is metaphorically illustrated as search in a foggy landscape, where one can't see the local peaks around you. Generally, studies in the NK field have assumed that the higher the K/N ratio, the more rugged landscape, and the more difficult it is for the agent to search in it [32]. This assumption is based on the idea that landscapes are either smooth or rugged and that no other features of a landscape interfere with the assessment of the difficulty of a landscape. However, a number of recent organizational studies have set forth to develop new types of landscapes. For example, Ethiraj and Levinthal [33] study the role of modular interdependence structure and Winter et al. [34] study search in a fractal landscape. More recently, Rahmandad [35] has proposed a variant of a fitness model (PN) where interdependence is modeled as an interaction polynomial, where both the sign and magnitude of the epistatic effect can be modeled. Our work builds on and extends these studies, in the sense that we propose that organizational studies consider and look into different ways of modeling complexity as well as other features, i.e. not just the K/N ratio.

Despite the fact that the notion of epistatic interactions, as outlined above, is widely used in e.g. organizational theory, its use in quantifying the difficulty of a problem has been criticized [36, 37] in particular due to the difficulty of identifying measures of epistasis that have adequate predictive power [7]. There are at least two main limitations to using epistasis measures as proxies for problem complexity. First, epistatic interactions can be both positive and negative. Whether an interaction effect between two alleles is positive or negative has a significant impact on the difficulty of a problem, but epistasis measures (e.g. epistasis variance or correlation) cannot capture this distinction [38]. Second, empirical evidence suggests that epistatic interactions can occur at several levels, i.e. there are hierarchical interdependence structures. This has consequences for the long-term dynamics of the system [39].

Paralleling the above investigations into the value of using epistasis measures as proxies for problem complexity, there has been considerable development when it comes to the study of fitness landscapes [7, 22, 40, 41]. This entails a focus on whether our categorization of fitness landscapes captures the features we can expect to encounter in the empirical world. Thus, rather than quantitatively characterizing problem difficulty (via ruggedness measures such as the K/N ratio) the aim is to categorize fitness landscapes into different types in order to determine the appropriate algorithm to use in a search problem [22]. This approach would allow for "a deeper understanding of a whole problem class" [7] rather than a specific problem instance. Current research thus seeks to identify relevant features that can describe a fitness landscape as well as having known properties with respect to problem-solving difficulty [42]. For example, Malan and Engelbrecht [42] point out two relevant features - beyond the difference between smooth and ruggedness - that also influence a searcher's ability to navigate the landscape and reach the optimal solution; neutrality and deceptiveness (Fig. 1).

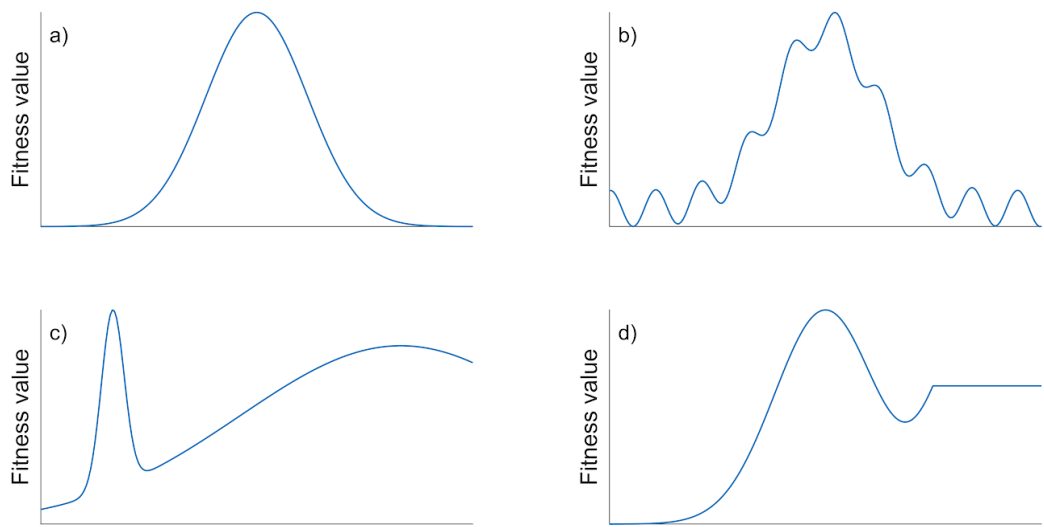

*Fig. 1 Landscape features adapted from Malan & Engelbrecht 2014. The landscapes are representations of different types of fitness functions where each position in the landscape has a fitness value. a) Smooth landscape. b) Rugged landscape. c) Deceptive landscape. d) Neutral Landscape*

In line with the prevalent approach in organizational theory, where the one-bit-flip hill-climbing algorithm is the dominant search behavior, in the following we describe how these two features (deceptiveness and neutrality) can affect the likelihood of finding the optimal solution for a classic one-bit-flip hill-climbing algorithm. The features' potential impact is not limited to this search heuristic.

Deceptiveness
Recent advances in biology point to the existence of higher-order epistatic interactions, which generate multidimensional landscapes [43, 44]. These interactions seem to be organized hierarchically in functional modules that interact with each other [44, 45]. This type of interaction structure is reminiscent of the hierarchical structure, which has been argued to be an essential feature of organizational problems, at least when it comes to innovation problems [21, 46,47]. In this context hierarchy is conceptualized as the composition of systems out of subsystems, where each subsystem, in turn, has its own hierarchy [48] until a certain level of fine-grained modularity is achieved. This is a qualitatively different kind of 'problem complexity' (as compared to landscape 'ruggedness') and the one most likely to be encountered in real-life problems [47-49]. In other words, the hierarchical decomposition and hierarchical interdependence are different from the one-level interdependence, which is captured by NK-like landscapes - see also Marengo et al. [50] for a more detailed account. Importantly, a K/N ratio can not capture if a problem is hierarchical.

Such hierarchical problems are likely to generate deceptive landscapes (cf. Fig. 1c where the majority of the landscape would steer a local-search hill-climber towards a local optimum, that is not the global optimum), according to Malan and Engelbrecht's [42] classification, since they generate so-called hierarchical traps where local-search gets stuck [49, 51, 52]. The interactions between building blocks make hierarchical problems deceptive (i.e. misleading according to Jones and Forrest [53]) in Hamming space at lower hierarchical levels, but fully non-deceptive at higher hierarchical levels [54]. Empirical studies in computer science analyzing how different computational algorithms solve various kinds of computer games also reveal that a measure of complexity does not capture the deceptiveness of the game [55].

## Neutrality

The metaphor of a rugged NK landscape focuses on the smooth vs. rugged distinction, and how to quantitatively measure the ruggedness. A different intuition about how evolutionary dynamics might be influenced by the underlying fitness function emerges from models that consider the possibility that some solutions have equal fitness. This was fueled by developments in molecular biology, which have questioned the 'rugged landscape' metaphor, in particular its explanation of speciation [56, 57]. This work was largely driven by the *neutral theory of molecular evolution* and in particular the observation that the majority of mutations at a molecular level do not affect the phenotype [18, 58]. The traditional NK framework assumes that once a population becomes stuck in a suboptimal peak it could only escape it if the fitness function was changed (e.g. shifting balance theory) or via a long jump. The neutral theory of molecular evolution relies on the conjunction that there must be a series of fitness neutral mutations that would allow organisms that were currently located in a suboptimal peak to 'escape', and undergo further evolution. Neutrality has also been observed in a quantum physics experiment, where it rendered sequences of 1D parameter optimizations unproductive [13].

A number of authors have introduced neutral extensions of the NK landscape and investigate how the new topology might influence the evolutionary processes [18, 56, 59, 60]. The implementations vary in both details and conclusions regarding the influence of neutrality on the features of the landscape [61], but they do conclusively show that neutrality is an important feature that influences search performance and is not captured by the traditional measures of ruggedness [7] commonly used in NK studies.

## Landscape ruggedness: modality and locality measures

We have addressed the pitfalls of using K or K/N ratios as measures of landscape ruggedness (the common approach in the social sciences) as well as identifying features that can impede any classifications of ruggedness. While it is clear that K influences how an agent is to search a given landscape, it is not clear how much epistasis "is needed to make a problem difficult" [62]. Thus, we present a number of alternative approaches to capture landscape ruggedness.

In computer science, a frequently used measure of landscape ruggedness is the number of local maxima or the modality of a landscape. The modality of a given landscape is often computed relative to the size of the fitness landscape: the higher the density of such local optima, the more complex the problem, i.e. the higher the likelihood that a solver will be stuck and unable to find the optimal solution. Note that the definition of a distance metric (and implicitly the neighborhood function) affects the number of local optima, since, by definition, for a problem $(X, f)$ and a neighborhood function $M$, a solution $x^*$ is called locally optimal with respect to $M$, if

$$f(x) \leq f(x^*) \text{ for all } x \in M(x) \quad (1)$$

Another perspective relies on the locality of a landscape, which is given by how closely together solutions with similar fitness values are located [63]. In general, the lower the distance, the higher the locality and better solutions are located closer together [7], thus arguably the easier it is to find a global optimum. One way of quantitatively measuring the locality of a landscape is a fitness distance correlation coefficient [53].

$$\rho_{FDC} = \frac{c_{fd}}{\sigma(f)\sigma(d_{opt})} \quad (2)$$

where

$$C_{fd} = \frac{1}{m}\sum_{i=1}^{m}(f_i - f)(d_{i,opt} - d_{opt}) \quad (3)$$

with $f$ as the mean value for the fitness function, $d_{\text{opt}}$ the mean value for the distance to the optimal solution, $f_i$ the fitness value for solution $i$ and $d_{i,\text{opt}}$ is the distance of solution $i$, to the optimal solution $x^*$. $\rho_{FDC}$ is thus the Pearson-correlation between the distance to the optimal solution and the fitness value.

The fitness-distance correlation coefficient, allows Jones and Forrest [53] to distinguish between three classes of landscapes:

Straightforward, for $\rho_{\text{FDC}} \leq -0.15$. This is the ideal case where the closer a solver gets to the global optimum, the higher the fitness. These cases are roughly correspondent to 'smooth' landscapes. NK problems where K ≤ 3, fall in this category.

Difficult $-0.15 < \rho_{\text{FDC}} < 0.15$ There is limited correlation between the fitness difference and the distance to the optimal solution. This makes such optimization problems very hard to solve and renders the search heuristics to random search. According to Jones and Forrest [53] as K increases over 3, NK landscapes quickly become uncorrelated and $\rho_{\text{FDC}}$ approaches 0. These are 'rugged' landscapes, with limited or uncorrelated ruggedness.

Misleading $\rho_{\text{FDC}} \geq 0.15$. There is an inverse correlation between the fitness difference and the distance to the optimal solution. Thus, the solver is drawn away from the global optimum. According to Malan and Engelbrecht's [40] classifications, these are 'deceptive landscapes'.

Jones & Forrest thus provide a quantitatively informed qualitative assessment of how difficult a landscape is to navigate in, which, however, requires an a priori knowledge of the global optimum. This is in line with a recent trend [40] arguing that we need to be able to classify landscapes and then identify which algorithm is appropriate, rather than searching for an absolute identification of the difficulty of a landscape (as K/N). In addition to providing information about the type of fitness landscape, the fitness distance correlation also manages to capture the challenge of deceptiveness, unlike the K/N ratio. However, Jones and Forrest [53] did not aim to capture neutrality, which still can influence the nature of the fitness landscape.

Thus, if neutrality is a feature that characterizes social science problems, caution should be used when characterizing the fitness landscape by relying on K/N ratios [63]. As Huynen et al. [64] argue, a small value for the fitness distance correlation (i.e. $-0.15 < \rho_{\text{FDC}} < 0.15$) that would normally be connected with a very rugged landscape, is not informative as to the ease or difficulty of finding the global optimum since local optima, when connected, are no longer local [64]. Lobo et al. [59] conclude that there is an interplay between the ruggedness and neutrality of the landscape, and their simulations suggest that the desirability of neutrality is contingent on the former. For instance, for rugged landscapes, neutrality is beneficial, while neutrality just makes adaptation slower in smooth landscapes.

In consequence, the quantitative ruggedness measures detailed in the previous section do not necessarily capture the relative ease or difficulty an adaptive solver would have on a landscape that has neutral ridges. In the following we move from an assessment of the nature of the task structure to the impact of search behaviors.

## Changing Search Strategies

Typically, an agent is set to primarily engage in one-bit-flips when searching the landscape, although other kinds of distant search strategies such as random search or search that match the underlying landscape have also been implemented [15, 26, 29]. We do not intend to discuss which search strategies should be applied in any given context, but want to highlight that the chosen search strategies (e.g. two-bit flips, rather than a one-bit-flip) not only reflect an assumption about search behavior but also constitute variations in distance metrics that influence the shape of the fitness landscape itself. In other words, the same fitness function can thus lead to qualitatively different landscapes, when assuming different kinds of search behavior [62]. To illustrate this point, Fig. 2 shows an example of the same function mapped onto three different landscapes using three different assumptions about search behavior. We used a dimensionality reduction method called t-SNE [65] that transforms high dimensional data to low-dimensional representations while approximately preserving pairwise similarities and global structure to create 3D visualizations of the multi-dimensional landscape. This visual transformation allows for a useful representation of relevant features of a landscape, such as how the connectedness of a landscape depends on search behavior assumptions. Given the fact that the t-SNE (t-distributed stochastic neighbor embedding) algorithm is stochastic, as is the NK fitness function, it should be noted that this is one possible illustration of one possible NK landscape with N=8, K=3. The illustration is not a general result for all NK landscapes of N=8 and K=3, and should be considered a visualization of a conceptual argument.

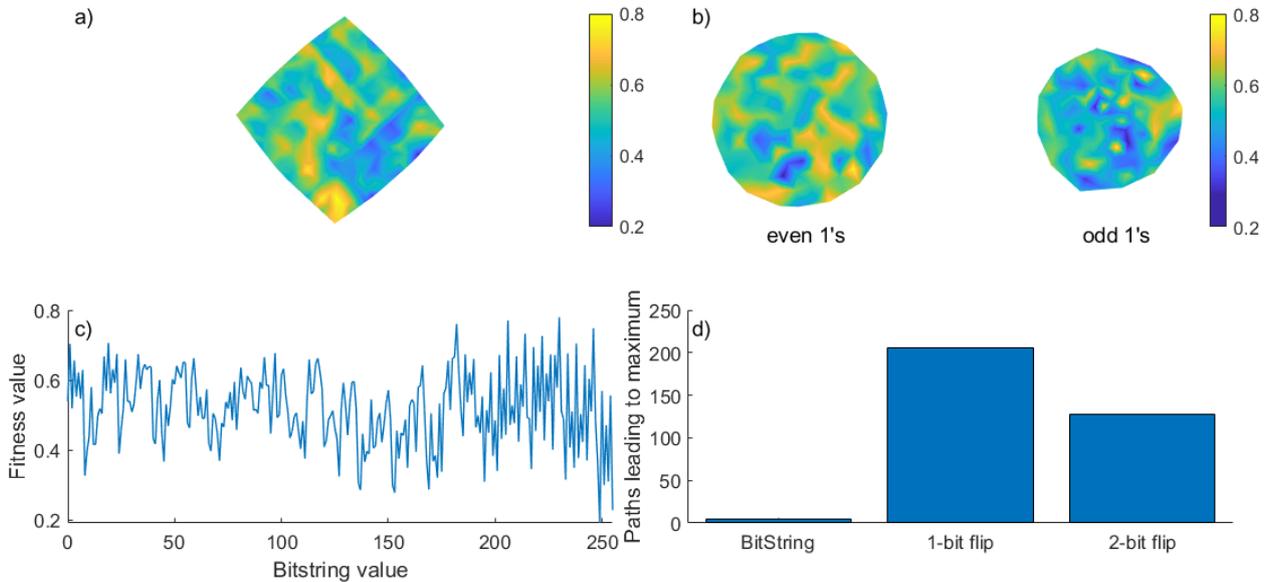

*Fig. 2 The same function (N=8, K=3) mapped with three different definitions of the neighborhood. a) NK landscape visualized by relying on a one-bit-flip, i.e. any two solutions are depicted next to each other if the Hamming distance between them is exactly one. b) NK landscape visualized by relying on a two-bit flip neighborhood structure, i.e. any two solutions are depicted next to each other if the Hamming distance between them is exactly two. The landscape split into two completely disconnected halves since any bitstring with an even number of 1's can never reach any of the bitstrings with an odd number of 1's using only two-bit flips, e.g. 0000000 can not be transformed into 00000001. c) NK function mapped by converting the bit of strings from a binary number to the decimal number (cf. Østman & Adami 2014) and assuming that solutions are similar if their decimal representation transformations are consecutive, i.e. 10011001 is transformed into 153 and its natural neighbors are 152 (10011000) and 154 (10011010). d) Comparative performance of c) bitstring, a) 1-bit flip and b) 2-bit flip algorithm.*

The visualization is informative in two ways. First, when assessing the visualization it is clear that the two-bit-flip search strategies (Fig. 2b) generate qualitatively different landscapes. Depending on the starting point, a subset of solutions is not connected in the graph. Similarly, if one attempts to traverse a sequence of consecutive numbers with increments of two, one generates two distinct and unconnected subsets: bit strings with an odd or even number of non-zero bits. Thus, the definition of the neighborhood function can effectively reduce (relative to the entire search space) the size of the landscape. Second, and more importantly, the heat-map of the landscape reveals information regarding the distribution of fitness scores. The blue (yellow) color reflects an area with poor (good) solutions, while the low-dimensional distance mapping approximates the distance in the high-dimensional space. If all yellow areas were directly connected it would indicate a smooth landscape, and the distribution of the colors (or heat) thus illustrates how rugged the landscape will appear to be for the given search strategy. The visualizations illustrate that the three neighborhood representations yield three different landscape topologies, i.e. smoother gradients (yellow areas are more connected) such as the even number of 1's side of the two-bit flip mean that it would be easier for an agent to find the global optimum, while 'patchier' surfaces translate into a lower likelihood of success, such as the one-bit-flip and the odd number of 1's side landscape generated by the two-bit flip. Equivalently, in the decimal representation, one can assess the difficulty of finding the global optimum (the highest fitness value), by looking at the shape of the generated curve (Fig. 2c). Since the decimal representation is arbitrary, but still systematic, the 'decimal' landscape is very 'rugged'; thus, a solver will likely be stuck in a suboptimal solution.

Overall, the visualizations demonstrate that a landscape is not merely inherently easy or difficult since the difficulty will depend on the agent's prevalent search strategy. The extent of ruggedness, as defined in Kauffman's original model, is assumed to be given by a one-bit-flip mutation of the candidate solution [6]. This was argued to be a reasonable assumption for how genes mutate and adapt.

Local-search is the default search behavior in these settings because, as highlighted by March [66], organizations suffer from cognitive myopia. Indeed empirical studies of the NK [e.g. 22,67] also make this assumption when investigating the topology of the fitness landscape. They find mixed results. Specifically, Fleming and Sorenson [67] argue that indeed inventors rely more on intelligent recombination and less on local-search. However, more recently Ganco [22] finds empirical evidence that supports the NK model with local-search *is* a useful model for innovation. While empirical work on organizational search is therefore mixed, as Frenken et al. [68] point out, the assumption of one-bit-flip is of limited relevance in the context of human search behaviors, since such a one-bit-flip conception does not fit human behavior: human problem-solvers are not constrained to engaging in small, incremental changes. Indeed, a number of experimental studies all showcase an average Hamming distance of about 2.5 [11, 32, 28]. In another landmark study Mason and Watts [69] find that empirical data of search behavior deviates significantly from current models of search. This seems to be in part due to the individual's ability to exploit local spatial correlations in the landscape [69]. Thus, while the myopic search argument due to cognitive limitations is appealing, it is ultimately the individuals not 'organizations' that engage in any kind of behavior (as microfoundations literature is trying to remind us). So ultimately, it appears problematic to consider one-bit-flips archetypical. The challenge outlined above is not limited to a non-hierarchical NK landscape, but can also be extended to a hierarchical search environment. Fig. 3 shows the visualization of a hierarchical problem using a one-bit-flip hill-climber (Fig. 3a) and a 'chunking' algorithm (Fig. 3b) that was tailored specifically for this problem (see Appendix 1 for a list of moves). The chunking algorithm was developed after analyzing quantitative and qualitative data obtained from humans attempting to solve the H-XOR problem [28]. Two of the authors of this manuscript have coded the moves made by human solvers. Upon obtaining an exhaustive list of 'possible moves' a chunking hill-climbing algorithm was developed around them. The moves consist of manipulations of chunks ('modules') that are manipulated both individually and dependent on other chunks.

We then generated maps based on agents starting from random points in the landscape. The underlying fitness function is identical, yet the visualization shows that the 'chunking' landscape turns out much smoother than the one-bit-flip landscape and even more so for the bitstring landscape. Since the H-XOR function has $2^{N/2}$ local optima for the one-bit-flip hill-climber (as illustrated by the two global peaks in Fig. 3c), the probability that a given point in the one-bit landscape is connected with a path to the global optimum is significantly lower as compared to the chunking algorithm (Fig. 3d). In other words, this problem is 'deceptive' for a one-bit-hill-climber but not for an agent relying on a problem representation that can exploit the problem structure. Again, a simple K/N ratio does not capture this type of feature of the landscape and it is not possible to disentangle assumptions about search behavior and the assessment of the fitness landscape characteristics.

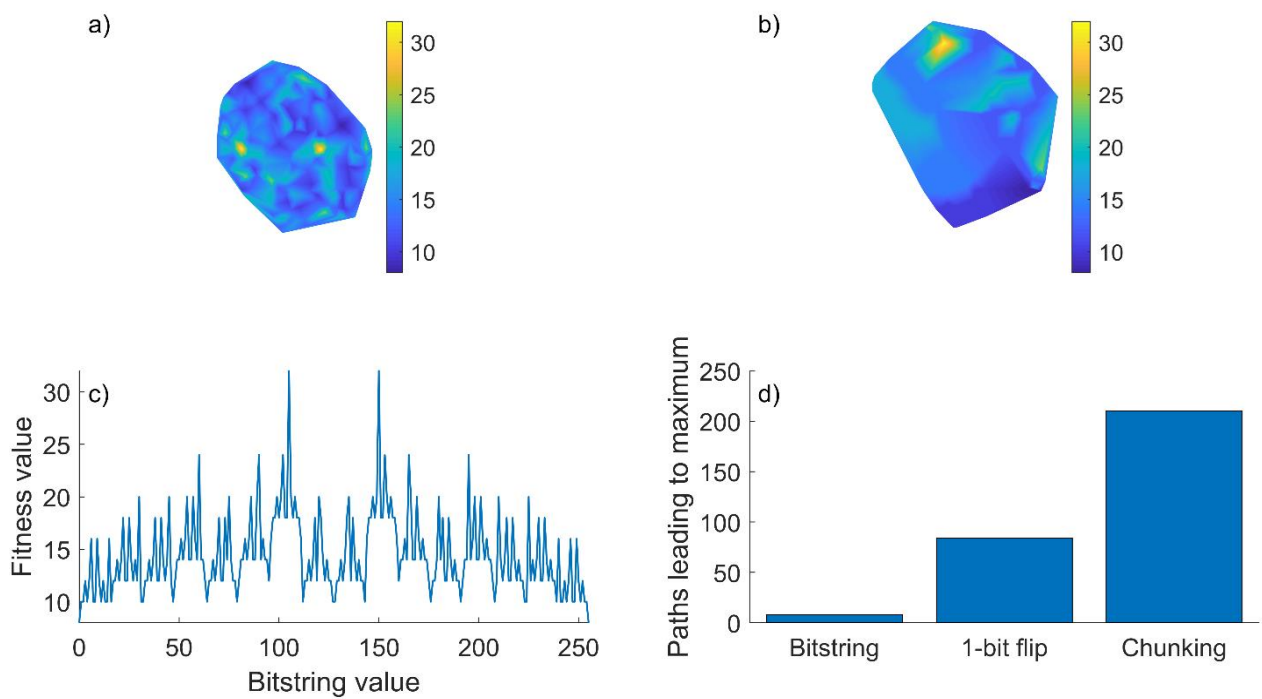

*Fig. 3 Same function (H-XOR, N=8) mapped with three different definitions of the neighborhood. a) 1-bit flip representation. b) Chunking algorithm representation. c) Bitstring representation in a decimal system (see Figure 2 for explanation of bitstrings). d) Comparative performance of bitstring (c), 1-bit flip (a) and chunking algorithm (b).*

## Discussion

While the simulation approach has gained attention in high-status outlets within the social sciences in general and organizational theory in particular, we here want to acknowledge and address the skeptical concerns still being raised about theoretical assumptions [70] and the weak empirical grounding of these assumptions [69,71,72]. Much like in the original biological setup, the organizational literature has mostly had simplistic assumptions about agent behavior, embedded in a relatively undefined fitness landscape [10, 73]. However, unlike microbiology, where evolutionary forces are well known [74] defining human search behaviors and assessing the ruggedness of the fitness landscape in this conceptual framework turns out to be elusive.

Based on visualizations and simulations, we showed that the neighborhood function influences how rugged a landscape will appear to the searcher. Since one can't just assume one-bit-flip search [28,32], this turns out to be a fundamental challenge to any a priori assessment of fitness landscapes. Second, even if one assumes one-bit-flip search, there are substantial challenges to traditional, quantitative measurements. Landscapes can be hierarchical and thus deceptive [55, 32], and neutrality can influence how adaptive search unfolds in a given fitness landscape [18].

Given our limited understanding about the genotype-phenotype mapping in a technological setting [75], we suggest that the focus should not be on the statistical features of the landscape to be searched under the one-bit-flip condition, but on how the interplay of search behaviors and the different natures of interdependence structures translate into problem-solving performance. One way forward, as suggested by current developments in computer science, is to revert to what Wright [4] and Kauffman [6] originally proposed; relying on fitness landscapes to first acquire a 'rough' image of a problem class, instead of investigating specific instances [7, 42]. This would entail moving beyond a simple, quantitative assessment of the topology (cf. the K/N ratio) or even the more sophisticated approach by Jones and Forrest [53].

These challenges illustrate that we need further empirical evidence on what relevant fitness landscapes actually look like, rather than assuming that a given K (that does not capture complexity) fits the empirical realm one is interested in. In biology, the empirical evidence towards the existence of multi-modal landscapes with numerous epistatic interactions continues to increase [41] and in cultural evolution there is also an ongoing discussion about the degree to which the biological world resembles NK landscapes [12]. Ganco [23] and Fleming and Sorenson [67] relied on patent analysis to empirically investigate if the NK model matches the fitness landscape of organizational outcomes. Yet, they reach different conclusions and we would argue that investigating if and when organizational landscapes might be deceptive or neutral would be a relevant avenue for future research

For example, neutrality can be argued to be generated by discretization. That is, while the contributions for the overall fitness can in principle be real numbers, and thus infinite many possibilities exist, one could argue that not all differences are of consequence and therefore at some level of sensitivity some local peaks are indistinguishable from the global peak [59]. Neutrality can also be generated by probabilistic draws, where some combinations are nonviable, and therefore the landscape would be less of a Mount Fuji and rather resemble a maze of narrow ridges traversing a landscape of treacherous valleys [57]. Both these distinct mechanisms can arguably be seen at play in different organizational settings. The case for the first one can be made since some organizational choices (e.g. hiring, promotions etc) do not have a visible and immediate impact on performance. However, as argued by [31], these neutral choices accumulate and do eventually have an impact on performance. On the other hand, the literature on entrepreneurship suggests that e.g. most software startups fail before reaching their commercial potential [76] so it would seem that when one attempts to model starting a business the best approximation is Gavrilet's [57] treacherous landscape.

Finally, deceptiveness is by definition given by having at least one local maximum with a larger basin of attraction then the global maximum. Indeed, in the case of hierarchical interdependence, complementary can be responsible for this[1]. The H-XOR function is an interesting illustration of this mechanism: when subcomponents can be optimized independently or quasi independently, this creates a landscape that can be characterized by deception, since there are multiple optimal ways of combining the lower level building blocks but awareness of the highest hierarchical level is required for finding the global optimum. As argued earlier, we believe this hierarchical interdependence to be an essential feature for organizations [50] and therefore some level of deceptiveness is unavoidable. In essence, such insights would allow us to make more sophisticated decisions about what kind of theoretical modeling assumptions to rely on, in which situations.

---

[1] We would like to thank and acknowledge an anonymous reviewer for pointing this out.

The NK model has been remarkably successful in progressing our ability to computationally model search challenges [9]. Yet, we argue that moving away from 'armchair speculations' [77] regarding human search behavior and the nature of the problem is paramount, as seemingly innocuous assumptions can drastically change the problem-solving performance. This requires empirical investigations into how empirical landscapes actually look like as well as acknowledging that any assessment of the fitness landscape necessarily needs to take the neighborhood function, i.e. search behavior, into consideration. This path should ideally lead to better simulations that inform our understanding of the empirical world and are further calibrated by empirical insights.

## Data Availability
The codes behind the simulations presented in the paper are publicly available through git https://gitlab.com/scienceathome-public/complexity-2020.git.

## Funding Statement
This work was partially supported by the Carlsberg Foundation [Grant id: CF16 – 0593] and the John Templeton Foundation [Grant id: 60969].

## Author Contributions
All authors participated in the initial conceptualization. O.V. and M.K.P did the formal analysis and wrote the main manuscript text and M.K.P prepared the figures. C.B. and J.F.S were in charge of project administration and supervision. All authors reviewed the manuscript.

## Competing interests
The authors declare no competing interests.

# Appendix: Operations for the 'chunking algorithm'

Chunks 8
Inverse all                              e.g. 01111111 → 10000000
Mirror all                                e.g. 01111111 → 11111110

Chunks 4 4
Inverse the 1st chunk                 e.g. 0111 1111 → 1000 1111
Inverse the 2nd chunk                e.g. 0111 1111 → 0111 0000

Mirror the 1st chunk                  e.g. 0111 1111 → 1110 1111
Mirror the 2nd chunk                 e.g. 0111 1110 → 0111 0111

Permute the 1st and 2nd chunk      e.g. 0111 1111 → 1111 0111

Chunks 3 2 3
Inverse the 1st chunk                 e.g. 011 11 111 → 100 11 111
Inverse the 2nd chunk                e.g. 011 11 111 → 011 00 111
Inverse the 3rd chunk                e.g. 011 11 111 → 011 11 000

Mirror the 1st chunk                  e.g. 011 11 111 → 110 11 000
Mirror the 2nd chunk                 e.g. 011 10 111 → 011 01 000
Mirror the 3rd chunk                 e.g. 011 11 011 → 011 11 110

Permute the 1st and 3rd chunk      e.g. 011 11 111 → 111 11 011

Chunks 2 2 2 2
Inverse the 1st chunk                 e.g. 01 11 11 11 → 10 11 11 11
Inverse the 2nd chunk                e.g. 01 11 11 11 → 01 00 11 11
Inverse the 3rd chunk                e.g. 01 11 11 11 → 01 11 00 11
Inverse the 4th chunk                e.g. 01 11 11 11 → 01 11 11 00

Mirror the 1st chunk                  e.g. 01 01 01 01 → 10 01 01 01
Mirror the 2nd chunk                 e.g. 01 01 01 01 → 01 10 01 01
Mirror the 3rd chunk                 e.g. 01 01 01 01 → 01 01 10 01
Mirror the 4th chunk                 e.g. 01 01 01 01 → 01 01 01 10

Permute the 1st and 2nd chunk      e.g. 01 11 11 11 → 11 01 11 11
Permute the 1st and 4th chunk      e.g. 01 11 11 11 → 11 11 11 01
Permute the 2nd and 3rd chunk     e.g. 11 01 11 11 → 11 11 01 11
Permute the 3rd and 4th chunk     e.g. 11 11 01 11 → 11 11 11 01